\documentclass[12pt]{article} 
\def\cdate{{February 16, 2024}}
\usepackage{amsfonts}
\usepackage{amsmath}
\usepackage{latexsym}
\usepackage{amssymb}
\usepackage{txfonts}
\usepackage{bm} 
\usepackage[american]{babel}
\usepackage{graphicx}
\usepackage{framed}
\usepackage{xcolor}
\definecolor{mygray}{gray}{0.95} 
\definecolor{mydarkgray}{gray}{0.70} 
\colorlet{shadecolor}{mygray}

\makeatletter
\def\timenow{%
\@tempcnta=\time \divide\@tempcnta by 60 \number\@tempcnta:\multiply
\@tempcnta by 60 \@tempcntb=\time \advance\@tempcntb by -\@tempcnta
\ifnum\@tempcntb <10 0\number\@tempcntb\else\number\@tempcntb\fi}
\newcounter{outputpage}
\renewcommand{\@oddhead}
{\stepcounter{outputpage}\hfill\hfill\theoutputpage}
\renewcommand{\@evenhead}
{\stepcounter{outputpage}\hfill\hfill\theoutputpage}
\renewcommand{\@oddfoot}
{\vbox{
\vspace{3pt}
\hfil
{\scriptsize\textit{
}}
\hfil
}}
\renewcommand{\@evenfoot}
{\vbox{
\vspace{3pt}
\hfil
{\scriptsize\textit{
}}
\hfil
}}
\makeatother

\def\nmt{{
\null
\vspace{-4cm}
\par
\hspace*{50truemm}{\hrulefill}
\par
\vskip-4truemm
\par
\hspace*{50truemm}{\hrulefill}
\par\vskip5mm
\par
\hspace*{50truemm}{{\large\sc 
New Mexico Tech {\rm 
(\cdate)
}}}\vskip4mm
\par
\hspace*{50truemm}{\hrulefill}
\par
\vskip-4truemm
\par
\hspace*{50truemm}{\hrulefill}
\par}}

\def\RR{\mathbb{R}} 
\def\CC{\mathbb{C}}

\def\cA{\mathcal{A}}

\def\cD{\mathcal{D}}
\def\cF{\mathcal{F}}
 
\def\cH{\mathcal{H}}

\def\cL{\mathcal{L}}

\def\cL{\mathcal{L}}
\def\cR{\mathcal{R}}

\def\const{\mathrm{const\,}}

\def\diag{\mathrm{diag\,}}

\def\sdet{\mathrm{sdet\,}}

\def\tr{\mathrm{tr\,}}

\def\la{\langle}
\def\ra{\rangle}
\def\proof{\par\noindent{\textbf{\textit{Proof:\;}}}}

\def\nn{{\nonumber}}
\def\be{\begin{equation}} 
\def\ee{\end{equation}} 
\def\bea{\begin{eqnarray}} 
\def\eea{\end{eqnarray}} 
\def\bed{\begin{definition}{\ }}
\def\eed{\end{definition}}
\def\ed{\end{document}}
\def\bp{\begin{proposition}}
\def\ep{\end{proposition}}
\def\bc{\begin{center}}
\def\ec{\end{center}}
\def\bi{\begin{itemize}} 
\def\ei{\end{itemize}} 
\def\benum{\begin{enumerate}} 
\def\eenum{\end{enumerate}} 
\def\bmp{\begin{minipage}} 
\def\emp{\end{minipage}} 

\newtheorem{lemma}{Lemma}

\newtheorem{proposition}{Proposition}

\newtheorem{definition}{Definition}
\begin{document}

\begin{titlepage}
\thispagestyle{empty}
\nmt

\bigskip
\bigskip
\bigskip
\bigskip
\centerline{\huge\textbf{MOND via Matrix Gravity}}
\bigskip
\bigskip
\bigskip
\centerline{\Large\bf Ivan Avramidi and Roberto Niardi}
\bigskip
\centerline{\it Department of Mathematics}
\centerline{\it New Mexico Institute of Mining and Technology}
\centerline{\it Socorro, NM 87801, USA}
\centerline{\it E-mail: ivan.avramidi@nmt.edu, roberto.niardi@student.nmt.edu}
\bigskip
\medskip

\begin{abstract}

MOND theory has arisen as a promising alternative to dark matter in explaining 
the collection of discrepancies that constitute the so-called missing mass problem. 
The MOND paradigm is briefly reviewed. It is shown that MOND theory can be 
incorporated in the framework of the recently proposed
Matrix Gravity. In particular, we demonstrate that 
Matrix Gravity contains MOND as a particular case, which adds to the validity of 
Matrix Gravity and proves it is deserving of further inquiry. 

\end{abstract}

\end{titlepage}


\section{Introduction}
\setcounter{equation}{0}

The theory of gravity, which Sir Isaac Newton famously introduced in the 17th
century and Albert Einstein's General Relativity later developed in the early
20th century, has served as a cornerstone of our knowledge of the fundamental
forces that shape the world for many years. Numerous phenomena, ranging from the
motion of celestial bodies to the behavior of matter on Earth, have been
successfully described by these theories. However, a rising body of evidence
indicates that adjustments to our existing theories of gravity may be required
as our scientific knowledge has grown and our observational abilities have
increased (for an extensive review, see \cite{famaey2012}).

Here we list some of the problems that constitute the so-called missing mass
problem:

\begin{itemize}

	\item In 1932, Oort \cite{Oort1932} found that the total number of visible
	stars in the solar neighborhood was insufficient to fully explain the vertical
	motions of stars in the Milky Way disk. The observed star vertical oscillations
	could not be explained entirely by the luminous matter. This experimental fact
	became known as Oort discrepancy.

	\item Concomitantly, Zwicky \cite{Zwicky1933} noted that in order for galaxies
	within galaxy clusters to form a bound system over a significant portion of
	cosmic time, their velocity dispersion should have been way lower than
	measured.

	\item Galactic disk stability was one of the earliest recent signs of the
	necessity for dark matter. Cold, self-gravitating disks were shown to be
	extremely unstable in early simulations \cite{OstrikerAndPeebles1973}. A
	potential well was necessary for the stability of the system. This could be
	provided by a halo of dark matter, as Ostriker \& Peebles
	\cite{OstrikerAndPeebles1973} proposed in 1973. This would prevent the growth
	of these instabilities, which would have destroyed the whole system on short a
	time scale, compared to a Hubble time.

	\item Bosma \cite{Bosma1981} and Rubin \cite{Rubin1982} established that the
	rotation curves of spiral galaxies are approximately flat: according to
	Newtonian gravity, when the majority of its mass is enclosed, a system should
	have a rotation curve that diminishes in a Keplerian fashion, similar to the
	solar system: $v \sim r^{-1/2}$.  Nonetheless, rotation curves for spiral
	galaxies tended to remain roughly flat with increasing radius instead: $v \sim
	{\rm constant}$. A second time, the proposal by Ostriker and Peebles of a dark
	matter halo would have been the solution to this mystery.

	\item Further anomalies come from the observation of dwarf spheroidal Milky Way
	satellite galaxies \cite{SimonAndGeha2007, Walker2009}. By analyzing the motion
	of individual stars that constitute these galaxies, the masses of these systems
	can be inferred, and it is way beyond the mass visible in luminous stars.

	\item Another piece of evidence comes from larger scales. It is the so called
	\lq\lq timing argument'' in the Local Group \cite{KahnAndWoltjer1959}. At this
	time, the Milky Way and Andromeda are approaching one another. But it is widely
	believed that the distance between the material constituting these galaxies was
	initially increasing, as a consequence of the Big Bang expansion. In order for
	the two to have overcome the initial expansion, a gravitating mass greater than
	their average mass has to be located between them. The mass-to-light ratio
	which is necessary to explain the approaching velocity currently measured at
	the present time is way beyond the mass-to-light ratio of the stars themselves,
	which have the same mass-to-light ratio of our Sun, as order of magnitude
	\cite{Bell2003}. \item Cluster of galaxies exhibit mass discrepancies, too. In
	particular, this is observed in three different ways, which give similar
	estimates for the amount of missing mass. The first way is through the
	measurement of the redshifts of individual cluster members; from these,
	velocity dispersions can be inferred and consequently mass-to-light ratio can
	be estimated \cite{Bahcall1995}. The second way is through mapping the
	temperature and emission of the diffuse intra-cluster gas, which emits in the
	X-rays; an estimate of the mass is obtained through the equation of hydrostatic
	equilibrium \cite{Giodini2009}. The third way is through the lensing produced
	by these cluster of galaxies \cite{Kneib1996}.  In all cases, the amount of
	mass needed to explain these phenomena is way beyond the amount actually
	observed.

	\item Further evidence comes from cosmology: by comparing the rate of growth of
	large scale structures and the estimate of the baryonic mass density 
$\Omega_{\rm baryonic}$
	coming from Big Bang Nucleosynthesis (BBN), we deduce that $\Omega_{\rm gravitating} >
	\Omega_{\rm baryonic}$ \cite{Davis2000, Walker1991, Copi1995}. Therefore, whatever
	dark matter is, it has to be non-baryonic.

	\item Another cosmological constraint comes from the Cosmic Microwave Background 
(CMB) radiation: from the temperature
	fluctuations at the time of decoupling between photons and baryons, we infer
	that the Universe was very homogeneous at the beginning, approximately to one
	in $10^5$. Nonetheless, the Universe today presents remarkable density
	contrasts between various astrophysical objects, and the emptiness of
	intergalactic space. Gravity is the only interaction we know that could be
	responsible for the formation of such structures. But the baryon density from
	BBN can only produce a growth factor of $\sim 10^2$ in a Hubble time
	\cite{Silk1967}, while the observed one is roughly $10^5$. In this context,
	dark matter would be the ideal candidate to enhance the growth rate. Moreover,
	this dark matter must not interact with electromagnetic radiation, because
	otherwise its contribution on the CMB could not be distinguished from the one
	produced by baryons.

\end{itemize}

All the experimental facts listed above (and many others) have prompted
scientists to think that those anomalies were caused by not taking into account
a new kind of matter with peculiar features: dark matter. Nonetheless, so far
dark matter has always escaped detection.

Therefore, another path for the explanation of these anomalies is admitting that
our understanding of gravity is still incomplete, and a modification of General
Relativity is necessary, even if General Relativity is still the most successful
and observationally tested theory to study acceleration as well as galaxy
clusters and large-scale structures. The Modified Newtonian Dynamics (MOND)
theory has emerged as a promising alternative framework that modifies the laws
of gravity at very small accelerations
in order to explain the observed dynamics without invoking the
presence of dark matter. It is worth noting that not all the observations in
this field have been fully explained using MOND so far.

This paper explores the incorporation of MOND theory into the framework of
an alternative theory of gravity, called 
Matrix Gravity, \cite{avramidi2003,avramidi2004a,avramidi2004b}. By introducing
noncommutativity into the geometry, it becomes possible to take into account
hypothetical additional physical degrees of freedom. 
We hope that this work will give some hint on the present
state of the Universe and other astrophysical phenomena.

We begin by reviewing the main concepts of MOND theory in Sec. 2
and its various relativistic versions in Sec. 3. 
In Sec. 4 we describe the non-commutative metric used as a main ingredient
of the Matrix Gravity and develop the corresponding noncommutative algebra.
In Sec. 5 we introduce the non-commutative connection and the corresponding
curvature. We analyze the advantages and disadvantages of various different 
approaches and choose the most consistent approach.
In Sec. 6 we use those noncommutative geometric objects to construct the corresponding
Lagrangian of Matrix Gravity.
In Sec. 7 we construct the action of a noncommutative Matrix Gravity and
explore possible ways in which the noncommutative framework can be used to reformulate
MOND, leading to a modified gravitational theory that incorporates
noncommutativity as a fundamental aspect of the metric tensor. 
We consider a very special case of Matrix Gravity when the noncommutative metric
is a two-dimensional diagonal matrix, which effectively leads to a bimetric theory.
We show that this model reproduces a generalization of the BIMOND theory proposed by
other authors \cite{milgrom2009,milgrom2010a,bekenstein2004}.
In Sec. 8 we analyze the static weak field limit of our model and show that it
effectively reproduces the QUMOND theory proposed by \cite{milgrom2010b}. 
In Sec. 9 we summarize our results and discuss the open problems.
We examine the
consequences of this modification for the behavior of gravity. The utility of
this work is twofold. One one hand, it validates the proposal for Matrix Gravity
by relating it to real experimental facts. On the other hand, it places the
phenomenological MOND theory on a more fundamental basis connecting it to a
solid mathematical foundation by preserving all symmetries of the model.

\section{MOND Theory}
\setcounter{equation}{0}

To fix notation, we work in the three-dimensional Euclidean space $\RR^3$ with
the standard Euclidean metric $\delta_{ij}$ and Cartesian coordinates $x^i$,
where the Latin indices are running over $i=1,2,3$. For the relativistic theory
we use the units with $c=1$ and the coordinates $(x^\mu)=(t,x^i)$, where the
Greek indices run over $\mu=0,1,2,3$ so that $x^0=t$; the Minkowski metric is
$\eta_{\mu\nu}=\diag(-1,1,1,1)$. The volume element is denoted by
$dx=dx^1\,dx^2\,dx^3$. Also, summation over repeated indices is assumed if not
specified otherwise. The partial derivatives with respect to $x^\mu$ are denoted
by $\partial_\mu$ and the Laplacian by $\Delta=\nabla_i\nabla^i$; the norm of a
vector $A_i$ is denoted by $|A|=\sqrt{A_i A^i}$.

The basic equation of the Newtonian gravity is the Poisson equation
\be
\Delta\varphi=4\pi G\rho,
\ee
where $\varphi$ is the gravitational potential,
$\rho$ is the mass density and $G$ is the gravitational constant,
which is obtained from the action
\be
S=S_{\rm Newton}+S_{\rm matter},
\ee
with
\bea
S_{\rm Newton} &=& -\frac{1}{8\pi G}\int dt\int\limits_{\Omega} dx\;|\nabla\varphi|^2,
\\[10pt]
S_{\rm matter} &=& -\int dt\int\limits_{\Omega} dx\;\rho\varphi,
\eea
where $\Omega$ is a region in space.

The idea of the Modified Newtonian Dynamics (MOND) theory of gravity suggested by Milgrom
\cite{milgrom1983} is to modify the Newtonian dynamics in the weak-acceleration regime,
$a<a_0$, with a universal acceleration constant
\be
a_0\approx 10^{-8} \;\mathrm{cm}\; \mathrm{sec}^{-2}.
\label{25fin}
\ee
This amounts to modifying the Poisson equation 
by \cite{bekenstein1984}
\be
\nabla_i\left[\mu\left(\frac{|\nabla\varphi|}{a_0}\right)\nabla_i\varphi\right]=4\pi G\rho,
\ee
where  $\mu(x)$ is a function that is supposed to have the properties
\be
\mu(x)
\sim 
\left\{
\begin{array}{ll}
1 & \mbox{ for } x>>1,
\\[5pt]
x & \mbox{ for } x<< 1.
\end{array}
\right.
\ee
This equation is obtained from the following action
\be
S_{\rm MOND}=-\frac{a_0^2}{8\pi G}\int dt\int\limits_\Omega dx\;
F\left(\frac{|\nabla\varphi|^2}{a^2_0}\right),
\ee
where the function $F(z)$ satisfies
\be
F(z)
\sim 
\left\{
\begin{array}{ll}
z & \mbox{ for } z>>1,
\\[12pt]
\frac{2}{3}z^{3/2} & \mbox{ for } z<< 1.
\end{array}
\right.
\ee
The function $\mu(x)$ is then determined by
$\mu(x)=F'(x^2)$.

An alternative modification called QUMOND \cite{milgrom2010b}
requires the introduction of two potential
functions
\be
S_{\rm QUMOND}=\frac{1}{8\pi G}\int dt\int\limits_\Omega dx\;
\left\{
-2\nabla_i\varphi\nabla_i\psi
+a_0^2Q\left(\frac{|\nabla\psi|^2}{a^2_0}\right)
\right\},
\ee
where the function $Q(z)$ satisfies
\be
Q(z)
\sim 
\left\{
\begin{array}{ll}
z & \mbox{ for } z>>1,
\\[12pt]
\frac{4}{3}z^{3/4} & \mbox{ for } z<< 1.
\end{array}
\right.
\ee
This leads to two equations
\bea
\Delta \psi &=& 4\pi G\rho,
\\
\Delta \varphi &=& \nabla_i\left[\nu\left(\frac{|\nabla\psi|}{a_0}\right)\nabla_i\psi\right],
\eea
where $\nu(x)=Q'(x^2)$.
By choosing new variable
\be
\phi=\varphi-\psi
\ee
this action takes the form
\be
S_{\rm QUMOND}=\frac{1}{8\pi G}\int dt\int\limits_\Omega dx\;
\left\{
-|\nabla\varphi|^2
+|\nabla\phi|^2
+a_0^2H\left(\frac{|\nabla(\varphi-\phi)|^2}{a^2_0}\right)
\right\},
\label{215fin}
\ee
where
\be
H(z)=Q(z)-z.
\ee

\section{Relativistic MOND Theories}
\setcounter{equation}{0}

Milgrom also proposed a bimetric relativistic MOND theory (called BIMOND)
\cite{milgrom2009,milgrom2010a} with two metrics, $g_{\mu\nu}$ and $\hat
g_{\mu\nu}$, and two types of matter, $\rho$ and $\hat\rho$ (called the twin
matter). Let $M$ be a region in spacetime with a spacelike boundary $\partial
M$. Let $x^\mu$ be the spacetime coordinates and $y^i$ be the coordinates on the
boundary. Let $\gamma_{ij}$ and $\hat\gamma_{ij}$ be the induced metrics on the
boundary. Let $g=-\det g_{\mu\nu}$,  $\hat g=-\det \hat g_{\mu\nu}$, and
$\gamma=\det \gamma_{ij}$,  $\hat \gamma=\det \hat \gamma_{ij}$. Let $R$ and
$\hat R$ be the scalar curvatures of the metrics $g_{\mu\nu}$ and $\hat
g_{\mu\nu}$, $K$ and $\hat K$ are the corresponding extrinsic curvatures of the
boundary. Let $\Gamma^\alpha{}_{\mu\beta}$ and $\hat \Gamma^\alpha{}_{\mu\beta}$
be the Christoffel symbols of the metrics $g_{\mu\nu}$ and $\hat g_{\mu\nu}$,
\be
C^\alpha{}_{\mu\beta}=\Gamma^\alpha{}_{\mu\beta}-\hat \Gamma^\alpha{}_{\mu\beta},
\ee
and
\be
X=g^{\mu\nu}\left(C^\alpha{}_{\mu\beta}C^\beta{}_{\nu\alpha}
-C^\alpha{}_{\mu\nu}C^\beta{}_{\beta\alpha}\right).
\ee
The action of BIMOND theory has the form
\bea
S_{\rm BIMOND} &=&
\frac{1}{16\pi G}
\int\limits_M dx
\left\{\beta g^{1/2}R+\alpha\hat g^{1/2}\hat R
-2 a_0^2g^{1/4}\hat g^{1/4}f\left(\frac{X}{a_0^2}\right)
\right\}
\nn\\
&&
+\frac{1}{8\pi G}
\int\limits_{\partial M} dy\;
\left(\beta\gamma^{1/2}K+\alpha\hat\gamma^{1/2}\hat K\right),
\label{215iga}
\eea
where 
$dx=dx^0\,dx^1\,dx^2\,dx^3$, $dy=dy^1\,dy^2\,dy^3$,
$\alpha$ and $\beta$ are some parameters,
and $f(x)$ is a function that has the properties
\be
f(x)
\sim 
\left\{
\begin{array}{ll}
\const & \mbox{ for } x>>1,
\\[12pt]
\frac{4}{3}x^{3/4} & \mbox{ for } x<< 1.
\end{array}
\right.
\label{220fin}
\ee

In the static weak field limit with $g_{0i}=0$ and
\bea
g_{00}=-1-2\varphi, 
\qquad
\hat g_{00}=-1-2\hat\varphi, 
\eea
we have
\be
X=|\nabla\psi|^2,
\ee
where
\be
\psi=\varphi-\hat\varphi.
\ee
Therefore, 
the field equations become
\bea
\beta\Delta\varphi &=& 4\pi G\rho
+\nabla_i\left[f'\left(\frac{|\nabla\psi|^2}{a_0^2}\right)\nabla_i\psi\right],
\\[10pt]
\alpha\Delta\hat\varphi &=& 4\pi G\hat\rho
-\nabla_i\left[f'\left(\frac{|\nabla\psi|^2}{a_0^2}\right)\nabla_i\psi\right].
\eea
By combining these equations we get a linear equation
\be
\Delta(\beta\varphi+\alpha\hat\varphi) = 4\pi G(\rho+\hat\rho);
\ee
In a particular case, $\beta=1$, $\alpha=-1$ this equation determines the function $\psi$
\be
\Delta \psi = 4\pi G(\rho+\hat\rho),
\ee
and the functions $\varphi$ and $\hat\varphi$ are determined by the above equations.
This is equivalent to non-relativistic QUMOND theory.

Bekenstein \cite{bekenstein2004} made another remarkable attempt at building a
relativistic version of MOND; in this theory, called TeVeS, the gravitational field is
described by three dynamical objects: a metric tensor $g_{\mu\nu}$, a normalized
time-like vector field $U_\mu$, and a scalar field $\phi$, after which the
theory is named. In TeVeS the physical metric $\tilde{g}_{\mu\nu}$ (the one
coupled with matter) is obtained by multiplying the Einstein metric (the one
whose Lagrangian is $g^{1/2} R$) by a factor $e^{-2\phi}$ in the spacetime
directions orthogonal to $U_\mu$, and by a factor $e^{2\phi}$ along $U_\mu$. It
can be expressed in the following way:
\begin{eqnarray}
\tilde{g}_{\mu\nu} = e^{-2\phi}{g}_{\mu\nu} - 2 U_\mu U_\nu \sinh(2\phi) \, .
\end{eqnarray}
It can be shown that TeVeS has well-behaved Newtonian and MOND limits, predicts
significant gravitational lensing and, for suitable initial data, does not
contain superluminal propagation of tensor, vector and scalar waves.

In this regard, Skordis and Zlosnik
\cite{SkordisAndZlosnik2019,SkordisAndZlosnik2021} worked on a previously
unknown class of tensor-vector-scalar theories in which the speed of the tensor
mode is always equal to the speed of light. Moreover, they were able to show
that the theory reproduces the observed cosmic microwave background and mass
power spectra on linear cosmological scales.

Following a different approach, Navarro and Von Acoleyen
\cite{navarroANDvanacoleyen2006} modified General Relativity in a MOND-like
fashion introducing in the Einstein action a correction which is logarithmic in
the curvature tensors. This produces, among other things, both a long-distance
modification of gravity and short-distance corrections to Newton's law.

\section{Non-commutative Metric}
\setcounter{equation}{0}

An extended theory of gravity called Matrix Gravity (MG) was developed in our
papers \cite{avramidi2003,avramidi2004a,avramidi2004b}. In this approach all
matter fields are supposed to be vector-valued in a complex $N$-dimensional
vector space and gravity is described by a $N\times N$ matrix-valued
self-adjoint symmetric 2-tensor field $a^{\mu\nu}$ on a four-dimensional
space-time manifold. We briefly review this approach (with some modifications)
and apply it to construct a phenomenological MOND model which is, in fact, a
generalization of Milgrom's BIMOND theory.

The main idea of this approach is to replace the ingredients
of Riemannian geometry, like the metric, connection and curvature,
by the real elements of a noncommutative complex associative algebra
$L(\cH)$ of operators in a Hilbert space $\cH$.
We define the symmetrized
multilinear product
of self-adjoint operators 
\be
\la\la A_1\cdots A_k\ra\ra = \frac{1}{k!}
\sum_{\varphi\in S_k}A_{\varphi(1)}\cdots A_{\varphi(k)},
\ee
where the summation goes over the permutation group $S_k$.

It is worth pointing out that this notation comes with a caveat. 
Consider two matrices $A$, $B$, and their product $C$. We have

\be
\la\la C \ra\ra = C \neq \la\la AB \ra\ra =  \frac{1}{2}
(AB + BA). 
\ee
This happens because the expression of the symmetrized product depends on the
number of factors, therefore the operations of taking the symmetrized product
and substituting a matrix with a product of matrices do not commute.

Obviously, the symmetrized product is completely symmetric, that is,
\be
\la\la A_1\cdots A_i\cdots A_j\cdots A_k\ra\ra = 
\la\la A_1\cdots A_j\cdots A_i\cdots A_k\ra\ra.
\ee

In particular, this defines a
 bilinear product
\be
A\circ B=\la\la AB\ra\ra=\frac{1}{2}(AB+BA);
\ee
it is easy to see that it is commutative but not associative.
Further, we define a linear functional (called the trace)
on the algebra,
\be
\tr: L(\cH)\to \RR,
\ee
which has the cyclic property
\be
\tr (A_1A_2\cdots A_k ) = \tr ( A_2\cdots A_kA_1).
\ee
Most of the time we will think of the 
Hilbert space $\cH$ as a finite-dimensional vector space $V$ and 
the algebra $L(\cH)$ simply as a matrix
algebra $GL(V,\CC)$. So, the trace will be just the usual trace of a 
finite-dimensional matrix.
Notice that the multilinear product satisfies the Leibnitz rule
\be
D\la\la A_1\cdots A_k\ra\ra
=\sum_{j=1}^n \la\la A_1 \cdots (DA_j)\cdots A_k\ra\ra,
\ee
where $D$ is a derivative.

Let $A^{\mu}{}_{\nu}$ be a matrix-valued tensor. It is easy to see that
the tensor 
\be
T_{\nu_1\dots\nu_n}=\varepsilon_{\mu_1\dots\mu_n}
\la\la A^{\mu_1}{}_{\nu_1}\cdots A^{\mu_n}{}_{\nu_n}\ra\ra,
\ee
where $\varepsilon_{\mu_1\dots\mu_n}$ is the Levi-Civita symbol,
is completely anti-symmetric and, therefore, it defines
a matrix that we call the symmetrized determinant

\be
\varepsilon_{\mu_1\dots\mu_n}
\la\la A^{\mu_1}{}_{\nu_1}\cdots A^{\mu_n}{}_{\nu_n}\ra\ra
=\varepsilon_{\nu_1\dots\nu_n}\; \sdet A^{\mu}{}_{\nu},
\ee
that is, 
\bea
\sdet A^{\mu}{}_{\nu} &=& \frac{1}{n!}\varepsilon_{\mu_1\dots\mu_n}
\varepsilon^{\nu_1\dots\nu_n}
\la\la A^{\mu_1}{}_{\nu_1}\cdots A^{\mu_n}{}_{\nu_n}\ra\ra
\nn\\[10pt]
&=&
\la\la A^{\mu_1}{}_{[\mu_1}\cdots A^{\mu_n}{}_{\mu_n]}\ra\ra \, .
\eea
This is proved by using the property
\be
\varepsilon_{\mu_1\dots\mu_n}\varepsilon^{\nu_1\dots\nu_n}
=n!\delta^{\nu_1}{}_{[\mu_1}\cdots\delta^{\nu_n}{}_{\mu_n]} \, .
\ee

Similarly we define
\begin{eqnarray}
&&\sdet (B^{\mu\nu})=\frac{1}{n!}\varepsilon_{\mu_1\dots\mu_n}\varepsilon_{\nu_1\dots\nu_n}
\la\la B^{\mu_1 \nu_1}\cdots B^{\mu_n \nu_n}\ra\ra \, ,
\\[10pt]
&&\sdet (C_{\mu\nu})=\frac{1}{n!}\varepsilon^{\mu_1\dots\mu_n}\varepsilon^{\nu_1\dots\nu_n}
\la\la C_{\mu_1 \nu_1}\cdots C_{\mu_n \nu_n}\ra\ra \, .
\end{eqnarray}

Notice that the symmetrized determinant of a $(1,1)$
tensor is invariant under
diffeomorphisms, while the determinant of tensors
of type $(2,0)$ and $(0,2)$ define scalar densities
of the weight $-2$ and $2$ correspondingly. 
Also, under the transformation
\be
A'^\mu{}_\nu =U A^\mu{}_\nu U^{-1}
\ee
it transforms 
accordingly
\be
\sdet A'^{\mu}{}_{\nu}=U(\sdet A^{\mu}{}_{\nu}) U^{-1} \, .
\ee

Further, it is easy to see that the symmetrized determinant is multiplicative
for any two commuting tensors,
that is,
if
\be
C_{\mu\nu}B^{\alpha\beta}=B^{\alpha\beta}C_{\mu\nu},
\ee 
then for a tensor $A^\mu{}_\nu=B^{\mu\alpha}C_{\alpha\nu}$ there holds
\be
\sdet A=\sdet B\; \sdet C.
\ee

\begin{lemma}
There holds
\bea
\sdet (I+A)
&=& I+\sum_{k=1}^n \la\la A^{\mu_1}{}_{[\mu_1}\cdots A^{\mu_k}{}_{\mu_k]}\ra\ra
\label{4193iga}
\\
&=&
I+A+\frac{1}{2}A^2
-\frac{1}{2}\la\la A^\mu{}_\nu A^{\nu}{}_{\mu}\ra\ra
+\cdots
+\sdet A,
\nn
\eea
where
\be
A=A^\mu{}_\mu \, .
\ee
\end{lemma}


Let $a^{\mu\nu}$ be a  
self-adjoint symmetric matrix valued tensor, that is,
\be
(a^{\mu\nu})^*=a^{\mu\nu},
\qquad 
a^{\mu\nu}=a^{\nu\mu}.
\ee
Let  $b_{\mu\nu}$ be another matrix-values tensor defined by
\be
a^{\mu\nu}b_{\nu\alpha} =b_{\alpha\nu} a^{\nu\mu} = \delta^\mu{}_\alpha I;
\ee
this tensor is neither self-adjoint nor symmetric but rather satisfies
\be
(b_{\mu\nu})^*=b_{\nu\mu}.
\ee

Let $g_{\mu\nu}$ be a pseudo-Riemannian metric and $h^{\mu\nu}$ be some 
self-adjoint symmetric matrix valued tensor defined by
We define
\bea
a^{\mu\nu} &=& g^{\mu\nu}I+h^{\mu\nu},
\eea
where $I$ is the unit matrix.

As usual, we raise the indices with the contravariant metric $g^{\mu\nu}$ and lower the indices
with the covariant metric $g_{\mu\nu}$.
Then the tensor $b_{\mu\nu}$ can be found in perturbation theory in the tensor $h$
\be
b_{\mu\nu}=g_{\mu\nu}I
+\sum_{k=1}^\infty (-1)^k h_{\mu}{}^{\beta_1}h_{\beta_1}{}^{\beta_2}
\cdots h_{\beta_{k-2}}{}^{\beta_{k-1}}h_{\beta_{k-1}\nu}.
\ee

We compute the symmetrized determinant of the tensor $a^{\mu\nu}$
\be
\sdet a^{\mu\nu}=-g^{-1}(I+Q),
\ee
where  $g=-\det g_{\mu\nu}$ and
\be
Q=\sum_{k=1}^n \la\la h^{\mu_1}{}_{[\mu_1}\cdots h^{\mu_k}{}_{\mu_k]}\ra\ra;
\ee
this enables us to define
a self-adjoint matrix-valued density $\rho$ of weight $1$
by
\be
\rho=(-\sdet a^{\mu\nu})^{-1/2}
=g^{1/2}\Phi,
\label{327fin}
\ee
where
\bea
\Phi&=&(I+Q)^{-1/2}
\nn\\
&=&
I+\sum_{k=1}^\infty (-1)^k\frac{(2k-1)!}{k!(k-1)!2^{2k-1} }Q^k.
\label{328fin}
\eea

\section{Non-commutative Connection and Curvature}
\setcounter{equation}{0}

Let $\nabla_\mu$ be the covariant derivative defined with respect to the Levi-Civita connection
of the metric $g_{\mu\nu}$ so that
\bea
\nabla_\alpha a^{\mu\nu} &=& \nabla_\alpha h^{\mu\nu}.
\eea
Let $\varphi^\mu$ and $\psi_\mu$ be vector-valued
tensors.
We define the non-commutative covariant derivative by
\bea
\cD_\mu\varphi^\alpha=\partial_\mu\varphi^\alpha
+\cA^\alpha{}_{\beta\mu}\varphi^\beta,
\\
\cD_\mu\psi_\alpha=\partial_\mu\psi_\alpha
-\cA^\beta{}_{\alpha\mu}\psi_\beta,
\eea
where $\cA^\alpha{}_{\beta\mu}$ is a matrix valued connection.
We impose the following conditions on the connection.

\benum
\item
For the operator $\cD_\mu$ to be well defined it should
transform as connection under the diffeomorphisms
$x'^\mu=x'^\mu(x)$,
\bea
\cA'^\varepsilon{}_{\gamma\delta}
&=&
A^{\varepsilon}{}_{\theta} B^{\xi}{}_\gamma  B^{\phi}{}_\delta 
\left(\cA^\theta{}_{\xi\phi}
-E^{\theta}{}_{\xi\phi}\right),
\eea
where
\bea
A^\mu{}_\alpha &=& \frac{\partial x'^\mu}{\partial x^\alpha},
\\[5pt]
B^\mu{}_\alpha &=& \frac{\partial x^\mu}{\partial x'^\alpha},
\\[5pt]
E^\alpha{}_{\kappa\sigma} &=&  B^{\alpha}{}_{\lambda} C^{\lambda}{}_{\kappa\sigma},
\\[5pt]
C^\mu{}_{\alpha\sigma} &=& \frac{\partial^2 x'^\mu}{\partial x^\sigma\partial x^\alpha}.
\eea
Therefore, the connection should have the form
\be
\cA^{\alpha}{}_{\mu\nu}=\Gamma^{\alpha}{}_{\mu\nu}I
+\theta^{\alpha}{}_{\mu\nu},
\ee
where $\theta^{\alpha}{}_{\mu\nu}$ is a matrix-valued tensor.
Then the non-commutative covariant derivative takes the form
\bea
\cD_\mu\varphi^\alpha=\nabla_\mu\varphi^\alpha
+\theta^\alpha{}_{\beta\mu}\varphi^\beta,
\\[5pt]
\cD_\mu\psi_\alpha=\nabla_\mu\psi_\alpha
-\theta^\beta{}_{\alpha\mu}\psi_\beta.
\eea

\item
Let $\la\;,\;\ra$ be the inner product in the vector space $\cH$. Then we require
\be
\la \cD_\mu\psi_\alpha,\varphi^\alpha\ra
+\la \psi_\alpha,\cD_\mu\varphi^\alpha\ra
=\partial_\mu\la \psi_\alpha,\varphi^\alpha\ra.
\ee
Then the connection must be self-adjoint,
\be
\cA^{\alpha}{}_{\mu\nu}^*=\cA^{\alpha}{}_{\mu\nu},
\ee
and, therefore, the tensor $\theta$ is also
self-adjoint
\be
\theta^{\alpha}{}_{\mu\nu}^*=\theta^{\alpha}{}_{\mu\nu}.
\ee

\item
We also require the connection to be
torsion-free
\be
\cA^{\alpha}{}_{\mu\nu}=\cA^{\alpha}{}_{\nu\mu},
\ee
therefore,
\be
\theta^{\alpha}{}_{\mu\nu}=\theta^{\alpha}{}_{\nu\mu}.
\ee

\eenum

Now we need to fix the connection, i.e. to relate it somehow
to the tensors $a^{\mu\nu}$ and $b_{\mu\nu}$ and their first derivatives.
The derivatives of the tensor $a^{\mu\nu}$ transform
under diffeomorphism $x'=x'(x)$ as
\be
\frac{\partial a'^{\mu\nu}}{\partial x'^\rho}
=A^\mu{}_\alpha A^\nu{}_\beta B^\sigma{}_\rho
\left\{
\frac{\partial a^{\alpha\beta}}{\partial x^\sigma}
+E^{\alpha}{}_{\kappa\sigma} a^{\kappa\beta}
+E^{\beta}{}_{\kappa\sigma } a^{\alpha\kappa}
\right\}.
\ee
To fix the connection we should impose an {\it invariant compatibility
condition}. We will impose the compatibility condition in the form
\be
W^{\alpha\beta}{}_\mu=0,
\ee
where 
\be
W^{\alpha\beta}{}_\mu=\partial_\mu a^{\alpha\beta}
+c_1{\cal A}^\alpha{}_{\lambda\mu}a^{\lambda\beta}
+c_2{\cal A}^\beta{}_{\lambda\mu}a^{\alpha\lambda}
+c_3 a^{\lambda\beta}{\cal A}^\alpha{}_{\lambda\mu}
+c_4 a^{\alpha\lambda}{\cal A}^\beta{}_{\lambda\mu}.
\ee
By using the transformation rule of the connection one can show that
this quantity transforms as a tensor if 
\be
c_1+c_3=1.\,
\qquad
c_2+c_4=1.
\ee
Then it can be written in the form
\bea
W^{\alpha\beta}{}_\mu &=& \nabla_\mu a^{\alpha\beta}
+c_1 \theta^\alpha{}_{\lambda\mu}a^{\lambda\beta}
+c_2 \theta^\beta{}_{\lambda\mu}a^{\alpha\lambda}
+(1-c_1) a^{\lambda\beta} \theta^\alpha{}_{\lambda\mu}
+(1-c_2) a^{\alpha\lambda}\theta^\beta{}_{\lambda\mu}
\nn\\[5pt]
&=&
\nabla_\mu a^{\alpha\beta}
+c_1 R^{\alpha\beta}{}_{\mu}
+c_2 R^{\beta\alpha}{}_{\mu}
+(1-c_1) L^{\alpha\beta}{}_{\mu}
+(1-c_2) L^{\beta\alpha}{}_{\mu},
\eea
where
\bea
R^{\alpha\beta}{}_\mu &=& \theta^{\alpha}{}_{\lambda\mu}a^{\beta\lambda},
\\
L^{\alpha\beta}{}_\mu &=& a^{\beta\lambda}\theta^{\alpha}{}_{\lambda\mu}.
\eea

Then the compatibility condition has the form
\be
\nabla_\mu a^{\alpha\beta}
+c_1 R^{\alpha\beta}{}_{\mu}
+c_2 R^{\beta\alpha}{}_{\mu}
+(1-c_1) L^{\alpha\beta}{}_{\mu}
+(1-c_2) L^{\beta\alpha}{}_{\mu}
=0.
\label{3122iga}
\ee
By antisymmetrizing over the indices $\alpha$ and $\beta$ we get
\be
(c_1-c_2)\left(R^{\alpha\beta}{}_{\mu}
-R^{\beta\alpha}{}_{\mu}
-L^{\alpha\beta}{}_{\mu}
+L^{\beta\alpha}{}_{\mu}
\right)
=0.
\ee
By taking the anti-adjoint part of eq. (\ref{3122iga}) we get
\bea
(2c_1-1) (R^{\alpha\beta}{}_{\mu}
-L^{\alpha\beta}{}_{\mu})
+(2c_2-1) (R^{\beta\alpha}{}_{\mu}
-L^{\beta\alpha}{}_{\mu})
=0.
\eea
We rewrite these equations in the form
\bea
(c_1-c_2)\left(S^{\alpha\beta}{}_{\mu}
-S^{\beta\alpha}{}_{\mu}
\right)
&=&0,
\\
(2c_1-1)S^{\alpha\beta}{}_{\mu}
+(2c_2-1)S^{\beta\alpha}{}_{\mu}
&=&0,
\eea
where
\be
S^{\alpha\beta}{}_\mu =
\frac{1}{2}
\left(L^{\alpha\beta}{}_\mu
-R^{\alpha\beta}{}_\mu\right)
=\frac{1}{2}[a^{\beta\lambda},\theta^{\alpha}{}_{\lambda\mu}].
\label{3112iga}
\ee
By decomposing the tensor $S$ in symmetric and anti-symmetric parts 
in the first two indices we get
\bea
(c_1-c_2)S^{[\alpha\beta]}{}_{\mu}
&=&0,
\\
(c_1+c_2-1)S^{(\alpha\beta)}{}_{\mu}
&=&0.
\eea
Therefore, if $c_1\ne c_2$ then the tensor $S$ is symmetric in the first two indices
\be
S^{[\alpha\beta]}{}_{\mu}=0,
\ee
and if $c_1+c_2\ne 1$ then the tensor $S$ is anti-symmetric in the first two indices
\be
S^{(\alpha\beta)}{}_\mu=0.
\ee
Thus, if both of these condistions are satisfied then the tensor $S$ vanishes,
\be
S^{\alpha\beta}{}_\mu=0.
\ee
These conditions puts severe algebraic constraints on the algebra. 
Therefore, we will 
choose $c_1$ and $c_2$ in such a way to avoid additional algebraic constraints
coming from the compatibility conditions. 


Let us try to use eq. (\ref{3122iga}) with $c_1=c_2=1$
or $c_1=c_2=0$.

{\it Case I:}
In the case
\be
c_1=c_2=1
\ee
the compatibility condition looks like
\be
\nabla_\mu a^{\alpha\beta}
+R^{\alpha\beta}{}_\mu
+R^{\beta\alpha}{}_\mu=0
\ee
or
\be
K^{\mu\alpha\beta}
+X^{\alpha\mu\beta}
+X^{\beta\mu\alpha}=0,
\ee
where
\bea
K^{\mu\alpha\beta} &=& a^{\mu\gamma}\nabla_\gamma a^{\alpha\beta},
\\[5pt]
X^{\alpha\beta\gamma} &=&  a^{\beta\lambda}\theta^{\alpha}{}_{\lambda\mu} a^{\mu\gamma}
\eea
This equation can be solved in a closed form.
The solution has the form
\be
X^{\alpha\mu\beta}
=-\frac{1}{2}\left(
K^{\alpha\mu\beta}
-K^{\mu\alpha\beta}
-K^{\beta\alpha\mu}
\right)
+Y^{\alpha\mu\beta},
\ee
where $Y$ is an arbitrary tensor which is antisymmetric in the first and the third index,
\be
Y^{\alpha\beta\mu}=-Y^{\mu\beta\alpha}.
\ee

{\it Case II:}
In the case 
\be
c_1=c_2=0
\ee
the compatibility condition looks like
\be
\nabla_\mu a^{\alpha\beta}
+L^{\alpha\beta}{}_\mu
+L^{\beta\alpha}{}_\mu=0
\ee
or
\be
N^{\mu\alpha\beta}
+X^{\alpha\mu\beta}
+X^{\beta\mu\alpha}=0,
\ee
where
\be
N^{\mu\alpha\beta}=(\nabla_\gamma a^{\alpha\beta})a^{\mu\gamma}.
\ee
This equation can be solved in a closed form.
The solution has the form
\be
X^{\alpha\mu\beta}
=-\frac{1}{2}\left(
N^{\alpha\mu\beta}
-N^{\mu\alpha\beta}
-N^{\beta\alpha\mu}
\right)
+Z^{\alpha\mu\beta},
\ee
where $Z$ is an arbitrary tensor which is antisymmetric in the first and the third index,
\be
Z^{\alpha\beta\mu}=-Z^{\mu\beta\alpha}
\ee
However, the tensors $K$ and $N$ are not self-adjoint and, therefore,
the connection is also not self-adjoint.


{\it Case III:}
We might try the compatibility condition with the noncommutative metric,
that is,
\bea
\cD_\mu (a^{\alpha\beta}\psi_\beta)
= a^{\alpha\beta}\cD_\mu\psi_\beta.
\eea
This condition is satisfied if
\be
\nabla_\mu a^{\alpha\beta}
+\theta^\alpha{}_{\lambda\mu}a^{\lambda\beta}
+a^{\alpha\lambda}\theta^\beta{}_{\lambda\mu}=0,
\label{3113iga}
\ee
which  is equivalent to 
\be
\nabla_\mu b_{\alpha\beta}
-\theta^\lambda{}_{\alpha\mu}b_{\lambda\beta}
-b_{\alpha\lambda}\theta^\lambda{}_{\beta\mu}=0,
\ee
and guarantees the compatibility condition
\bea
\cD_\mu (b_{\alpha\beta}\varphi^\beta)
= b_{\alpha\beta}\cD_\mu\varphi^\beta.
\eea
This compatibility condition
(\ref{3113iga}) corresponds to 
\be
c_1=1, \qquad
c_2=0.
\ee

In this case the compatibility conditions reads
\be
\nabla_\mu a^{\alpha\beta}
+R^{\alpha\beta}{}_\mu
+L^{\beta\alpha}{}_\mu=0,
\ee
and it cannot be solved in a closed form.
Let
\bea
B^{\alpha\beta}{}_\mu &=&
\frac{1}{2}
\left(L^{\alpha\beta}{}_\mu
+R^{\alpha\beta}{}_\mu\right)
=a^{\beta\lambda}\circ\theta^{\alpha}{}_{\lambda\mu},
\label{3111iga}
\eea
Then we have
\bea
B^{(\alpha\beta)}{}_\mu &=& -\frac{1}{2}\nabla_\mu a^{\alpha\beta},
\\[5pt]
S^{[\alpha\beta]}{}_{\mu} &=& 0,
\eea
which means
\be
[a^{\beta\lambda},\theta^\alpha{}_{\lambda\mu}]
=[a^{\alpha\lambda},\theta^\beta{}_{\lambda\mu}].
\ee
These constraints cannot be satisfied even in perturbation theory
for a generic matrix algebra.


{\it Case IV:}
That is why, we choose
\be
c_1=c_2=\frac{1}{2}.
\ee
In this case
\be
W^{\alpha\beta}{}_\mu=\partial_\mu a^{\alpha\beta}
+{\cal A}^\alpha{}_{\lambda\mu}\circ a^{\lambda\beta}
+{\cal A}^\beta{}_{\lambda\mu}\circ a^{\alpha\lambda}
\ee
and the compatibility condition takes the form
\bea
\nabla_\mu a^{\alpha\beta}
+\theta^{\alpha}{}_{\lambda\mu}\circ a^{\lambda\beta}
+\theta^{\beta}{}_{\lambda\mu}\circ a^{\alpha\lambda}=0.
\label{461fin}
\eea
Therefore, the compatibility condition fixes the symmetric part of the
tensor $B$,
\be
B^{(\alpha\beta)}{}_\mu
=-\frac{1}{2}\nabla_\mu a^{\alpha\beta},
\ee
but leaves the antisymmetric part of the tensor $B$ and the tensor $S$ arbitrary.
Therefore, there are no extra algebraic constraints. 

We solve the compatibility condition (\ref{461fin})
in perturbation theory
(indices are raised with the metric $g$).
We have
\be
\theta^{\alpha\beta}{}_{\mu}
+\theta^{\beta\alpha}{}_{\mu}
=-\nabla_\mu h^{\alpha\beta}
-\theta^{\alpha}{}_{\lambda\mu}\circ h^{\beta\lambda}
-\theta^{\beta}{}_{\lambda\mu}\circ h^{\alpha\lambda},
\ee
or
\be
\theta_{\alpha\beta\mu}
+\theta_{\beta\alpha\mu}
=-P_{\alpha\beta\mu},
\label{464fin}
\ee
where
\be
P_{\alpha\beta\mu}=\nabla_\mu h_{\alpha\beta}
+\theta_{\alpha\lambda\mu}\circ h^{\lambda}{}_{\beta}
+\theta_{\beta\lambda\mu}\circ h^{\lambda}{}_{\alpha}.
\ee
Recall that $\theta_{\mu\nu\alpha}$ is 
symmetric in the last two indices.

\begin{lemma}
Let $T_{\alpha\beta\mu}$ be a tensor satisfying the conditions
\bea
T_{\alpha\beta\mu} &=& T_{\alpha\mu\beta},
\\
T_{\alpha\beta\mu} + T_{\beta\alpha\mu}&=& V_{\alpha\beta\mu}.
\eea
Then 
\be
T_{\alpha\beta\mu}=\frac{1}{2}
\left(V_{\alpha\beta\mu}
+V_{\alpha\mu\beta}
-V_{\beta\mu\alpha}\right)
\ee

\end{lemma}
\proof
We have
\bea
T_{\alpha\beta\mu} + T_{\beta\alpha\mu}&=& V_{\alpha\beta\mu},
\\
T_{\mu\alpha\beta} + T_{\alpha\mu\beta}&=& V_{\alpha\mu\beta},
\\
-T_{\mu\beta\alpha} - T_{\beta\mu\alpha}&=& -V_{\beta\mu\alpha}.
\eea
By adding these equations we obtain the result.

The equation (\ref{464fin})
can be solved in terms of the tensor $P$,
\be
\theta_{\alpha\beta\mu}=\frac{1}{2}
\left(
P_{\beta\mu\alpha}
-P_{\alpha\mu\beta}
-P_{\alpha\beta\mu}\right),
\ee
that is,
\bea
\theta_{\alpha\beta\mu}&=&\frac{1}{2}
\Biggl\{
\nabla_\alpha h_{\mu\beta}
-\nabla_\mu h_{\alpha\beta}
-\nabla_\beta h_{\alpha\mu}
\\
&&
+h^{\lambda}{}_{\beta}\circ(\theta_{\mu\lambda\alpha}
-\theta_{\alpha\lambda\mu})
+h^{\lambda}{}_{\mu}\circ(\theta_{\beta\lambda\alpha}
-\theta_{\alpha\lambda\beta})
-h^{\lambda}{}_{\alpha}\circ(\theta_{\mu\lambda\beta}
+\theta_{\beta\lambda\mu})
\Biggr\}.
\nn
\eea

This equation
can be solved in perturbation theory.
By rescaling $h\mapsto \varkappa h$ and expanding the tensor $\theta$
in the power series
\be
\theta=\sum_{k=1}^\infty \varkappa^k \theta_{(k)},
\ee
we obtain in the first order
\be
\theta_{(1),\alpha\beta\mu}
=\frac{1}{2}
\left(
\nabla_\alpha h_{\mu\beta}
-\nabla_\mu h_{\alpha\beta}
-\nabla_\beta h_{\alpha\mu}
\right),
\ee
and for the higher orders, $k\ge 2$, we get the recurrence relations
\bea
\theta_{(k)\alpha\beta\mu}&=&
\frac{1}{2}
\Biggl\{
h^{\lambda}{}_{\beta}\circ(\theta_{(k-1)\mu\lambda\alpha}
-\theta_{\alpha\lambda\mu})
+h^{\lambda}{}_{\mu}\circ(\theta_{(k-1)\beta\lambda\alpha}
-\theta_{\alpha\lambda\beta})
\nn\\
&&
-h^{\lambda}{}_{\alpha}\circ(\theta_{(k-1)\mu\lambda\beta}
+\theta_{(k-1)\beta\lambda\mu})
\Biggr\}.
\eea
In particular, in the second order we get
\be
\theta_{(2),\alpha\beta\mu}
=\frac{1}{2}\Biggl\{
h^{\lambda}{}_{\alpha}\circ\nabla_{\lambda} h_{\mu\beta}
+h^{\lambda}{}_{\beta}\circ(\nabla_{\mu} h_{\alpha\lambda}
-\nabla_{\alpha} h_{\mu\lambda})
+h^{\lambda}{}_{\mu}\circ(\nabla_{\beta} h_{\alpha\lambda}
-\nabla_{\alpha} h_{\beta\lambda})
\Biggr\}.
\ee
It is important to emphasize that this defines the tensor $\theta$ uniquely.

We compute the contraction of the connection
\be
\cA^\alpha{}_{\alpha\mu}
=-\omega_{,\mu}+\theta_\mu,
\ee
where $\omega=-\log g^{1/2}$, and 
\be
\theta_\mu=\theta^\alpha{}_{\alpha\mu}.
\ee
In the first order we have simply
\be
\theta_{(1),\mu}=-\frac{1}{2}\nabla_\mu h,
\ee
where
\be
h=h^\mu{}_\mu.
\ee
For the second order we obtain
\be
\theta_{(2),\mu}=
\frac{1}{2}
h^{\alpha\beta}\circ\nabla_{\mu} h_{\alpha\beta}
=\frac{1}{4}\nabla_\mu(h^{\alpha\beta}\circ h_{\alpha\beta})
\ee

Similarly, we expand the symmetrized determinant,
$\rho=(-\sdet a^{\mu\nu})^{-1/2}$, of the noncommutative metric
more precisely, the matrix $\Phi=g^{-1/2}\rho$ defined by eqs.
(\ref{327fin}), (\ref{328fin}),
\be
\Phi=I+\sum_{k=1}^\infty 
\varkappa^k \Phi_{(k)}.
\ee
By using (\ref{328fin}) we get
\bea
\Phi_1 &=& -\frac{1}{2} h,
\\[10pt]
\Phi_2 &=& 
\frac{1}{8}h\circ h
+\frac{1}{4}h_{\alpha\beta}\circ h^{\alpha\beta}.
\eea
Therefore,
\be
\nabla_\mu\Phi_1=\theta_{(1),\mu}=-\frac{1}{2}\nabla_\mu h
\ee
and
\bea
\nabla_\mu\Phi_2 
&=& 
\frac{1}{4}h\circ\nabla_\mu h
+\frac{1}{4}\nabla_\mu(h_{\alpha\beta}\circ h^{\alpha\beta})
\nn\\[10pt]
&=&
\theta_{(1),\mu}\circ\Phi_1+\theta_{(2),\mu}.
\eea
Therefore, the matrix $\Phi$ satisfies the equation 
\be
\nabla_\mu\Phi-\theta_\mu\circ\Phi=O(\varkappa^3)
\ee
up to the third order.


Now we can define the curvature (noncommutative Riemann tensor) as follows
\bea
\cR^{\alpha}{}_{\beta\mu\nu}
&=&\partial_\mu \cA^\alpha{}_{\beta\nu}
-\partial_\nu \cA^\alpha{}_{\beta\mu}
+\cA^\alpha{}_{\gamma\mu}\circ\cA^\gamma{}_{\beta\nu}
-\cA^\alpha{}_{\gamma\nu}\circ\cA^\gamma{}_{\beta\mu}
\nn\\[5pt]
&=&
R^{\alpha}{}_{\beta\mu\nu}I
+\nabla_\mu \theta^\alpha{}_{\beta\nu}
-\nabla_\nu \theta^\alpha{}_{\beta\mu}
+\theta^\alpha{}_{\gamma\mu}\circ\theta^\gamma{}_{\beta\nu}
-\theta^\alpha{}_{\gamma\nu}\circ\theta^\gamma{}_{\beta\mu}.
\eea
This tensor is antisymmetric in the last two indices
and is self-adjoint. It naturally leads to the 
unique noncommutative Ricci tensor
\bea
\cR_{\beta\nu}=
\cR^{\mu}{}_{\beta\mu\nu}
&=&\partial_\mu \cA^\mu{}_{\beta\nu}
-\partial_\nu \cA_{\beta}
+\cA_{\gamma}\circ\cA^\gamma{}_{\beta\nu}
-\cA^\mu{}_{\gamma\nu}\circ\cA^\gamma{}_{\beta\mu}
\nn\\[5pt]
&=&
R_{\beta\nu}I
+\nabla_\mu \theta^\mu{}_{\beta\nu}
-\nabla_\nu \theta_{\beta}
+\theta_{\gamma}\circ\theta^\gamma{}_{\beta\nu}
-\theta^\mu{}_{\gamma\nu}\circ\theta^\gamma{}_{\beta\mu}.
\eea
This tensor is self-adjoint but not symmetric.
Next, we define the scalar curvature by
\bea
\cR &=& 
a^{\beta\nu}\circ\cR_{\beta\nu}
\nn\\[5pt]
&=&
a^{\beta\nu}\circ (R_{\beta\nu}I
+\nabla_\mu \theta^\mu{}_{\beta\nu}
-\nabla_\nu \theta_{\beta})
+\Theta,
\eea
where
\be
\Theta=
a^{\beta\nu}\circ\left(\theta_{\gamma}\circ\theta^\gamma{}_{\nu\beta}
-\theta^\mu{}_{\gamma\nu}\circ\theta^\gamma{}_{\mu\beta}
\right);
\label{492fin}
\ee
which is also self-adjoint.


\section{Noncommutative Lagrangian}
\setcounter{equation}{0}


To construct the action of the noncommutative gravity
we need to construct invariant scalar densities from
the noncommutative metric and its derivatives.
We can use the scalar curvature to define a
real scalar density 
\be
\cL = \frac{1}{N}\tr\left(\rho\; \cR\right);
\ee
more explicitly, this invariant has the form
\be
\cL =
\frac{1}{N} g^{1/2}\tr
(\Phi \circ a^{\beta\nu})\Biggl\{
R_{\beta\nu}I
+\nabla_\mu \theta^\mu{}_{\beta\nu}
-\nabla_\nu \theta_{\beta}
+ \theta_{\gamma}\circ\theta^\gamma{}_{\beta\nu}
-\theta^\mu{}_{\gamma\nu}\circ\theta^\gamma{}_{\beta\mu}
\Biggr\}.
\ee


We note that the commutators of the connection 
\be
W^{\alpha\beta}{}_{\mu\nu\rho\sigma}=
[\cA^\alpha{}_{\mu\nu},\cA^\beta{}_{\rho\sigma}]
=[\theta^\alpha{}_{\mu\nu},\theta^\beta{}_{\rho\sigma}]
\ee
as well as the commutators of the connection with the 
noncommutative metric $a^{\mu\nu}$,
\be
V^{\mu\nu\alpha}{}_{\rho\sigma}=[a^{\mu\nu},\cA^\alpha{}_{\rho\sigma}]
=[h^{\mu\nu},\theta^\alpha{}_{\rho\sigma}],
\ee
transform as tensors. These tensors are anti-self-adjoint.
One can use these tensors to construct real scalars; one needs to add
terms linear in $W$ and quadratic in $V$
\be
\cL\sim  \frac{1}{N}g^{1/2}\tr \Phi\Biggl\{
ia^{\mu\nu} W^{\alpha\beta}{}_{\mu\nu\alpha\beta}+\dots
+b_{\alpha\beta} V^{\mu\nu\alpha}{}_{\rho\sigma}V^{\rho\sigma\beta}{}_{\mu\nu}
+\dots
\Biggr\}
\ee

In particular, we define
a natural self-adjoint symmetric tensor is
\be
\cF_{\mu\nu}=iW^{\alpha\beta}{}_{\mu\nu\alpha\beta}
=i[\cA^\alpha{}_{\mu\nu},\cA_{\alpha}]
=i[\theta^\alpha{}_{\mu\nu},\theta_{\alpha}];
\ee
it defines the anti-self-adjoint  scalar
\be
\cF=ia^{\mu\nu}\circ[\theta^\alpha{}_{\mu\nu},\theta_{\alpha}],
\ee
which
can be used to form a real non-commutative action term
\be
\cL =\frac{1}{N}\tr\left(\rho\cF\right);
\ee
more explicitly
\be
\cL =\frac{1}{N}g^{1/2}\tr (\Phi\circ a^{\mu\nu})i[\theta^\alpha{}_{\mu\nu},\theta_{\alpha}].
\ee

We could also use double commutators to construct tensors
\be
[\cA^\lambda{}_{\phi\chi},[\cA^\alpha{}_{\mu\nu},\cA^\beta{}_{\rho\sigma}]]
=[\theta^\lambda{}_{\phi\chi},[\theta^\alpha{}_{\mu\nu},\theta^\beta{}_{\rho\sigma}]]
\ee
However, these tensors are of odd rank and it is impossible to construct
scalars out of them. Also, they are cubic in the derivatives, $(\nabla h)^3$.

\section{MOND-like Noncommutative Gravity}
\setcounter{equation}{0}

We assume that all matter tensor fields are vector valued and the covariant 
derivatives are defined with the noncommutative connection 
$\cA^\alpha{}_{\mu\nu}=\Gamma^\alpha{}_{\mu\nu}I+\theta^\alpha{}_{\mu\nu}$ 
(which is defined
in terms of the noncommutative metric $a^{\mu\nu}=g^{\mu\nu}I+h^{\mu\nu}$).
So, the action of the matter fields $\psi$ is generalized as follows
\be
S_{\rm mat}(\psi, g)\mapsto S_{\rm mat}(\psi, a)
=S_{\rm mat}(\psi, g, h).
\ee
The noncommutative MOND gravitational action is defined
by
\be
S_{\rm ncgrav}(a)=\frac{1}{16\pi G }\int\limits_M dx \;
\frac{1}{N}\tr\rho
\Biggl\{
\cR+\lambda\cF
-2a^2_0 f\left(\frac{\Theta}{a_0^2}\right)
\Biggr\},
\label{62fin}
\ee
where $G$ is the Newton constant, $\lambda$ is a coupling constant and
$f(x)$ is the MOND function (\ref{220fin}).
This function is supposed to have the following properties
\be
f(x)\sim 
\left\{
\begin{array}{ll}
\displaystyle\frac{\Lambda}{a^2_0}  & \mbox{ for } x>>1,
\\[15pt]
\displaystyle\frac{4}{3}x^{3/4} & \mbox{ for } x<<1,
\end{array}
\right.
\ee
where $\Lambda$ is the cosmological constant and $a_0$ is the MOND
parameter (\ref{25fin}).
This is a generalization of the bimetric MOND theory (BIMOND)
(\ref{215iga}).
Originally, the Matrix Gravity was proposed in
\cite{avramidi2003,avramidi2004a,avramidi2004b} without the function $f$, and,
that is why, it was called Matrix General Relativity. By adding the MOND
function $f$ to the Lagrangian the model becomes reminiscent of $f(R)$
generalizations of the single-metric General Relativity. The difference is
though that the matrix $\Theta$ is not the curvature but rather the product of
connections. Notice that for large connections, $\Theta >> a_0^2$, the function
$f$ is constant and, therefore, the model describes the Matrix General
Relativity. However, for small connections, that is,  $\Theta << a_0^2$, the
dynamics is rather different, more like the MOND theory.

In this paper we consider a very special
Abelian case, when the matrix $a^{\mu\nu}$ is two-dimensional, $N=2$, and
diagonal, that is,
\bea
a^{\mu\nu} &=&g^{\mu\nu}I+h^{\mu\nu}T,
\eea
where $g^{\mu\nu}$ is the (pseudo)-Riemannian metric,
$h^{\mu\nu}$ is a symmetric tensor and $T$ is the diagonal matrix
\be
T=
\left(
\begin{array}{cc}
1 & 0\\
0 & -1 \\
\end{array}
\right).
\ee
This can be rewritten in the form
\bea
a^{\mu\nu}
&=& g^{\mu\nu}_+P_++g^{\mu\nu}_-P_-,
\eea
where $g^{\mu\nu}_\pm$ are the symmetric tensors defined by
\be
g^{\mu\nu}_+=g^{\mu\nu}+h^{\mu\nu},
\qquad
g^{\mu\nu}_-=g^{\mu\nu}-h^{\mu\nu},
\ee
and
$P_+, P_-$ are the projections
\be
P_+=
\left(
\begin{array}{cc}
1 & 0\\
0 & 0 \\
\end{array}
\right),
\qquad
P_-=
\left(
\begin{array}{cc}
0 & 0\\
0 & 1 \\
\end{array}
\right).
\ee

In this case all geometric quantities diagonalize
\bea
b_{\mu\nu} &=& g^+_{\mu\nu}P_++g^-_{\mu\nu}P_-,
\\[5pt]
\rho &=&  g^{1/2}_+P_++g^{1/2}_-P_-,
\\[5pt]
\cA^\alpha{}_{\mu\nu}
&=&
\Gamma_+^\alpha{}_{\mu\nu}P_+
+\Gamma_{-}^\alpha{}_{\mu\nu}P_-,
\\[5pt]
\cR^\alpha{}_{\beta\mu\nu}
&=& R_+^\alpha{}_{\beta\mu\nu}P_+
+R_{-}^\alpha{}_{\beta\mu\nu}P_-.
\eea
The metric $g^{\mu\nu}$ is related to the tensors $g^{\mu\nu}_\pm$ by
\be
g^{\mu\nu}=\frac{1}{2}(g^{\mu\nu}_++g^{\mu\nu}_-),
\ee
and the tensor $\theta^\alpha{}_{\mu\nu}$ is then defined by
\be
\theta^\alpha{}_{\mu\nu} 
=\theta_{+}^\alpha{}_{\mu\nu}P_+
+\theta_{-}^\alpha{}_{\mu\nu}P_-,
\ee
where
\be
\theta_{\pm}^\alpha{}_{\mu\nu}
=\Gamma_\pm^\alpha{}_{\mu\nu}-\Gamma^\alpha{}_{\mu\nu},
\ee
$\Gamma^\alpha{}_{\mu\nu}$ are Christoffel symbols for the metric
$g_{\mu\nu}$ and $\Gamma^\alpha_\pm{}_{\mu\nu}$ are Christoffel symbols for the metric
$g^\pm_{\mu\nu}$.
Therefore, we obtain the quantity (\ref{492fin})
\be
\Theta=\Theta_+P_++\Theta_-P_-,
\ee
where
\be
\Theta_\pm
=g_\pm^{\beta\nu}\left(\theta^\pm_{\gamma}\theta_\pm^\gamma{}_{\nu\beta}
-\theta_\pm^\mu{}_{\gamma\nu}\theta_\pm^\gamma{}_{\mu\beta}
\right).
\ee

Therefore, the action (\ref{62fin}) becomes
\bea
S_{\rm ncgrav}(a) &=& \frac{1}{16\pi G }\int\limits_M dx \;
\frac{1}{2}\Biggl\{
g_+^{1/2}R_+
+g^{1/2}_-R_-
\nn\\
&&
-2a^2_0 g_+^{1/2}f\left(\frac{\Theta_+}{a_0^2}\right)
-2a^2_0 g_-^{1/2}f\left(\frac{\Theta_-}{a_0^2}\right)
\Biggr\},
\eea
where $g_{\pm}=-\det g^\pm_{\mu\nu}$ and $R_\pm$ are the scalar curvatures of
the metrics $g^\pm_{\mu\nu}$. It is clear that in this special case, the number
of degrees of freedom is doubled, compared to General Relativity. This is a
generalization of the relativistic bimetric BIMOND action (\ref{215iga}) that
should be carefully studied.
Of course, this model is just a bimetric theory
describing in the quadratic approximation two non-interacting massless spin $2$ particles;
with $4=2+2$ degrees of freedom.

\section{Static Weak Field Limit}
\setcounter{equation}{0}

In the static weak field limit we have
\bea
g^{\mu\nu}_\pm &=& \eta^{\mu\nu}+2\varphi_\pm\delta^{\mu\nu},
\\
g^{\mu\nu} &=& \eta^{\mu\nu}+2\varphi\delta^{\mu\nu}, 
\eea
where $\eta^{\mu\nu}=\diag(-1,1,1,1)$ is the Minkowski metric, and
\be
\varphi=\frac{1}{2}(\varphi_++\varphi_-),
\ee
so that, up to the second order in $\varphi$
\be
g_\pm^{1/2}=1-2\varphi_\pm.
\ee
For each metric we compute
the Christoffel symbols
in the first order
\bea
\Gamma^0{}_{00} &=& \Gamma^0{}_{ij} = \Gamma^i{}_{j0} = 0,
\\[5pt]
\Gamma^0{}_{0i} &=& \Gamma^i{}_{00} =
\varphi_{,i},
\\[5pt]
\Gamma^i{}_{jk} &=&
\varphi_{,i} \delta_{jk}
-\varphi_{,j}\delta_{ik}
-\varphi_{,k}\delta_{ij}.
\eea
Here, as usual, comma  denotes the partial derivative, $\varphi_{,i}=\partial_i\varphi$. 

Next, we compute the tensor $\theta^\mu{}_{\alpha\beta}$, in the first order,
\bea
\theta_\pm^0{}_{00} &=&\theta_\pm^0{}_{ij}=\theta_\pm^i{}_{j0}=0,
\\[5pt]
\theta_\pm^0{}_{0i} &=& \theta_+^i{}_{00} =
\pm\frac{1}{2}\psi_{,i},
\\[5pt]
\theta_\pm^i{}_{jk} &=&
\pm\frac{1}{2}
\left(\psi_{,i} \delta_{jk}
-\psi_{,j}\delta_{ik}
-\psi_{,k}\delta_{ij}
\right),
\eea
where
\be
\psi=\varphi_+-\varphi_-.
\ee
Then we compute the tensor $\theta_\mu$; we get
\bea
\theta^\pm_0 &=& 0,
\\
\theta^\pm_i &=& \mp \frac{3}{2}\psi_{,i}.
\eea
Finally, we compute the scalars $\Theta_\pm$ in the second order,
\be
\Theta_\pm=\frac{1}{2}\psi_{,i}\psi_{,i}
=\frac{1}{2}|\nabla\psi|^2.
\ee
We also need the Einstein-Hilbert action in the second order
(after integration by parts)
\be
\int dx\; g^{1/2}R = 
-\int dx\;2|\nabla \varphi|^2
\ee

By using the above formulas we obtain
the static weak field action
\be
S_{\rm ncgrav}(a)=-\frac{1}{8\pi G }\int\limits_M dx \;
\Biggl\{
|\nabla\varphi_+|^2
+|\nabla\varphi_-|^2
+a^2_0f\left(\frac{|\nabla\psi|^2}{2a_0^2}\right)
\Biggr\},
\ee
which is very similar to the QUMOND model (\ref{215fin}).
Assuming the matter action
\be
S_{\rm matter}=\int\limits_M dx\; \left\{
-\rho_+\varphi_+-\rho_-\varphi_-
\right\},
\ee
we obtain two coupled equations of motion
\bea
\Delta\varphi_+ &=& 4\pi G\rho_+
-\frac{1}{2}\nabla_i\left[f'\left(\frac{|\nabla\psi|^2}{2a_0^2}\right)\nabla_i\psi\right],
\\[5pt]
\Delta\varphi_- &=& 4\pi G\rho_-
+\frac{1}{2}\nabla_i\left[f'\left(\frac{|\nabla\psi|^2}{2a_0^2}\right)\nabla_i\psi\right].
\eea
By subtracting and summing these equations we get
\bea
\Delta\varphi &=& 4\pi G\rho,
\\
\Delta\psi &=& 4\pi G(\rho_+-\rho_-)
-\nabla_i\left[f'\left(\frac{|\nabla\psi|^2}{2a_0^2}\right)\nabla_i\psi\right],
\eea
where
\be
\rho=\frac{1}{2}(\rho_++\rho_-).
\ee
For a symmetric matter, $\rho_+=\rho_-=\rho$, these equations coincide with the QUMOND model
(\ref{215fin}).

\section{Discussion}
\setcounter{equation}{0}


It is apparent that our theory (\ref{62fin}) has a larger number of degrees of freedom than
General Relativity. The spectrum of Matrix Gravity was analyzed in
\cite{fucci09}. We would like to emphasize that $g^{\mu\nu}$ is a {\it fixed
non-dynamical auxiliary tensor} used to to parametrize the tensor
$a^{\mu\nu}=g^{\mu\nu}I+h^{\mu\nu}$. It can be chosen rather {\it arbitrarily}
(in particular, equal to the flat Minkowski metric) with the primary goal to
develop the geometric machinery. The main {\it dynamical variable is the
matrix-valued tensor $h^{\mu\nu}$}; in particular, it is not necessarily
traceless. Recall that a self-adjoint $N\times N$ matrix has $N^2$ independent
real components and a symmetric $2$-tensor has $10$ independent components in
$4$ dimensions. Therefore, one can decompose the tensor $h^{\mu\nu}$ as
\be
h^{\mu\nu}=h^{\mu\nu}_0 I+\sum_{j=1}^{N^2-1}h^{\mu\nu}_j iT_j,
\ee
where $T_j$ are the anti-self-adjoint generators of the group $SU(N)$,
and $h^{\mu\nu}_0$, $h^{\mu\nu}_j$, $j=1,\dots,N^2-1$ are some real
two-tensors. 
Therefore, the model contains $N^2$ symmetric two-tensor fields.

These tensors can be decomposed further in the canonical way
\be
h_{\mu\nu}=\bar h^{\perp}_{\mu\nu}+\frac{1}{4}g_{\mu\nu}\phi+2\nabla_{(\mu}\xi_{\nu)},
\ee
where the tensor field $\bar h^\perp_{\mu\nu}$ is transverse traceless,
\be
\nabla_\mu \bar h^{\perp \mu\nu}=0,
\qquad
g_{\mu\nu}\bar h^{\perp \mu\nu}=0.
\ee
Note that the scalar $\phi$ has one component, the vector $\xi_\mu$ has $4$
independent components and the transverse-traceless symmetric tensor $\bar
h^\perp_{\mu\nu}$ has $5$ components, giving the total of $10=5+1+4$ components.
In the standard regime (when the MOND function $f$ is constant) the quadratic
part of the action is similar to that in the standard General Relativity, the
only difference being that the fields are matrix-valued. Therefore, it describes
$N^2$ identical actions for each field $h^{\mu\nu}_0$, $h^{\mu\nu}_i$. Because
of the diffeomorphism invariance the vector field $\xi_\mu$ is non-physical and
drops out, leaving $6$ degrees of freedom. Of course, for massless spin $2$
particles, the $6$ components of $\bar h^{\perp \mu\nu}$ describe $2$ 
physical degrees of freedom corresponding to two polarizations of the graviton.
It would be interesting to do the canonical analysis of the MOND regime as well;
we might do this in a future work.


Let us summarize briefly our main goal and the results.
General Relativity, despite being a milestone in our understanding of the
Universe and its large structures, is flawed by the so-called missing mass
problem, which has been briefly described in the introduction. Dark matter has
been proposed as the solution for the explanation of these anomalies.
Nonetheless, it has so far always escaped detection. Another approach to dealing
with this issue is to modify our current theory of gravity, which is,
proposing modified versions of General Relativity. Among these attempts, 
non-relativistic MOND theory and its relativistic versions
 propose an elegant solution to the missing mass problem via a suitable
modification of the action. Such a modification is obtained by introducing an
interpolation function whose behavior is only known in two asymptotic regimes.

On the other hand, Matrix Gravity was developed totally independent and with
completely different goals, mainly by analogy with non-Abelian gauge theories
such as Quantum Chromodynamics. This theory has a richer structure than 
General Relativity and does not involve any phenomenological interpolating
functions such the MOND function. It was not proposed as an attempt to solve the
missing mass problem. The current paper is just an attempt to apply the 
approach of Matrix Gravity to the dark matter problem.
Our main (non-trivial) result is that the
non-commutative framework of Matrix Gravity can be used to reformulate MOND. In
particular, MOND appears to be a special case of Matrix Gravity in which the
non-commutative metric tensor is two-dimensional and diagonal. Further attention
and investigation are needed in order to understand the full geometrical and
dynamical structure of Matrix Gravity. In particular, the next step is tackling
the more general case in which there are off-diagonal terms and/or the rank of
the non-commutative metric tensor is greater than two. We hope that such a study
may give some hint of the present state of the Universe.


\end{document}